\documentclass{ws-mpla}

\begin{document}

\markboth{Wo-Lung Lee}
{Quantum Noises and the Large Scale Structure}

\catchline{}{}{}{}{}

\title{QUANTUM NOISES AND THE LARGE SCALE STRUCTURE
}

\author{Wo-Lung Lee}

\address{Department of Physics, National Taiwan Normal University \\
Taipei, Taiwan 116, Republic of China
\\
leewl@phy.ntnu.edu.tw}

\maketitle

\pub{Received (Day Month Year)}{Revised (Day Month Year)}

\begin{abstract}
We propose that cosmological density perturbation may originate from passive fluctuations of the inflaton, which are induced by colored quantum noise due to the coupling of the inflaton to the quantum environment. At small scales, the fluctuations grow with time to become nearly scale-invariant. However, the larger-scale modes cross out the horizon earlier and do not have enough time to grow, thus resulting in a suppression of the density perturbation. This may explain the observed low quadrupole in the CMB anisotropy data and and potentially unveil the initial time of inflation. We also discuss the implications to the running spectral index and the non-Gaussianity of the primordial density perturbation.

\end{abstract}

\ccode{PACS Nos.: 98.80.Cq, 98.70.Vc, 98.80.Es}

\section{Introduction}\label{aba:sec1}
The recently released data of the Wilkinson Microwave Anisotropy Probe 
(WMAP) confirmed the earlier COBE-DMR's observation about the deficiency in 
fluctuation power at the largest angular scales~\cite{cobe,wmap}. The amount 
of the quadrupole mode of the CMB temperature fluctuations is 
anomalously low if compared to the prediction of the $\Lambda$CDM model. The usual explanation to this low quadruple moment is to invoke the so-call ``cosmic variance", i.e. at large scales the CMB experiments are limited by the fact that we only have one sky to measure and so cannot pin down the cosmic average to infinite precision no matter how good the experiment is. Given the large uncertainties due to the cosmic variance, we might never know whether the phenomenon of the low CMB quadrupole constitutes a truly significant deviation from the standard cosmological expectations. 

Though the cosmic variance implies that we simply live in a universe with a low quadrupole moment for no special reason, the lack of power on the largest scale can be treated as a physical effect that demands a judicious explanation. There exists various viable mechanisms for generating this large-scale anomaly. So far, all the efforts can be classified into three essential categories: (1) a possible non-trivial topology(for example, see J.-P. Luminet {\it et al.}~\cite{topo} and the references therein); (2) a cut-off due to causality (for example, see A. Berera {\it et al.}~\cite{casul} and the references therein); (3) initial hybrid fluctuations (for example, see Lee \& Fang~\cite{hic}). Recently the colored noise has been considered to explain the anomaly in the context of stochastic inflation~\cite{mata}. Instead of a Heaviside window function used in the stochastic approach to inflation~\cite{star}, Ref.[6] adopts a Gaussian window function with a width characterizing the size of the coarse-grained domain~\cite{wini} which is then arbitrarily chosen to be comparable to the Hubble radius. This smooth window function eventually yields a blue tilt of the power spectrum on large scales which can be tuned to fit the WMAP large-scale anisotropy data. Nevertheless, in this classical approach they inevitably resort to
an {\it ad hoc} smoothing window function and therefore the width of the window function remains undetermined.

From the field theoretical point of view, on the other hand, interactions between the inflaton and other quantum fields in the very early universe may be inevitable. Such interactions essentially convert the potential energy stored in the inflaton field into radiation during~\cite{winf} or after inflation~\cite{linde}. Consequently, the evolution of inflationary density perturbations will be altered depending upon the details of the mechanisms~\cite{hic,noise}. Therefore, the interactions between the inflaton and other fields have not only provided us an opportunity to grasp the underlying physics of the inflation, but also explored the possibility of generating primordial density fluctuations with statistical properties deviated from the scale-invariance and the Gaussianity. 

\section{Power Spectrum Derived from the Noise-Driven Fluctuations}
To mimic the quantum environment, we consider a slow-rolling
inflaton $\phi$ coupled to a quantum massive scalar field
$\sigma$, with a Lagrangian given by~\cite{cnoise}
\begin{equation}
\mathcal{L}=\frac{1}{2}g^{\mu\nu}\partial_{\mu}\phi\,\partial_{\nu}\phi+
            \frac{1}{2}g^{\mu\nu}\partial_{\mu}\sigma\,\partial_{\nu}\sigma
             -V(\phi)-\frac{m_{\sigma}^{2}}{2}\sigma^{2}-
             \frac{g^{2}}{2}\phi^{2}\sigma^{2},
\label{model}
\end{equation}
where $V(\phi)$ is the inflaton potential that complies with the slow-roll
conditions and $g$ is a coupling constant. Thus, we can approximate the
space-time during inflation by a de Sitter metric given by
\begin{equation}
ds^{2}=a(\eta)^{2}(d\eta^{2}-d\vec{x}^{2}),
\end{equation}
where $\eta$ is the conformal time and $a(\eta)= -1/(H\eta)$ with
$H$ being the Hubble parameter. Here we rescale $a=1$ at the
initial time of the inflation era, $\eta_i= -1/H$. Following the
influence functional approach~\cite{noise,fey}, we trace out
$\sigma$ up to the one-loop level and thus obtain the
semiclassical Langevin equation for $\phi$, given by
\begin{eqnarray}
&&\ddot{\phi}+2aH\dot{\phi}-\nabla^{2}\phi+a^{2}\left(V'(\phi)
+g^{2}\langle\sigma^{2}\rangle\phi\right)-g^{4}a^{2}{\phi}\,\times
\nonumber \\
&&\int d^{4}x' a^4(\eta') \theta(\eta-\eta')\,i\, G_{-}(x,x')
{\phi}^{2}(x')=\frac{{\phi}}{a^{2}}\xi,
\label{lange}
\end{eqnarray}
where the dot and prime denote respectively differentiation with
respect to $\eta$ and $\phi$. The effects from the quantum field
on the inflaton are given by the dissipation via the kernel
$G_{-}$ as well as a stochastic force induced by the
multiplicative colored noise $\xi$ with
\begin{equation}
\langle\xi(x)\xi(x')\rangle= g^4 a^4(\eta) a^4(\eta') G_{+}(x,x').
\label{noise}
\end{equation}
The kernels $G_{\pm}$ in Eqs.~(\ref{lange}) and (\ref{noise}) can be
constructed from the Green's function of the $\sigma$ field
with respect to a particular choice of the initial vacuum state
which we will specify later, given by
\begin{equation}
G_{\pm}(x,x')=\langle\sigma(x)\sigma(x')\rangle^2 \pm
                               \langle\sigma(x')\sigma(x)\rangle^2.
\label{grfct}
\end{equation}

To solve Eq.~(\ref{lange}), let us first drop the dissipative term
which we will discuss later. Then, after
decomposing $\phi$ into a mean field and a classical perturbation:
$\phi(\eta,\vec x)={\bar\phi}(\eta) + {\varphi}(\eta,\vec x)$, we obtain
the linearized Langevin equation,
\begin{equation}
\ddot{\varphi}+2aH\dot{\varphi}-\nabla^{2}{\varphi}+
a^{2} m_{\varphi{\rm eff}}^2 {\varphi} = {{\bar\phi}\over a^2}\xi,
\label{varphieq}
\end{equation}
where the effective mass is
$m_{\varphi{\rm eff}}^2=V''({\bar\phi})+g^{2}\langle\sigma^{2}\rangle$
and the time evolution of $\bar\phi$ is governed by $V({\bar\phi})$.
The equation of motion for $\sigma$ from which we construct its Green's
function can be read off from its quadratic terms in the
Lagrangian~(\ref{model}) as
\begin{equation}
\ddot{\sigma}+2aH\dot{\sigma}-\nabla^{2}{\sigma}+ a^{2}
m_{\sigma{\rm eff}}^2 {\sigma} = 0,
\label{sigmaeq}
\end{equation}
where $m_{\sigma{\rm eff}}^2=m_{\sigma}^2+g^{2}{\bar\phi}^2$. 

The solution to Eq.~(\ref{varphieq}) is obtained~\cite{cnoise} as
\begin{equation}
\varphi_{\vec k}= -i\int_{\eta_i}^{\eta} d\eta'{\bar\phi}(\eta')
\xi_{\vec k}(\eta')\left[\varphi_k^{1}(\eta')\varphi_k^{2}(\eta)
                   - \varphi_k^{2}(\eta')\varphi_k^{1}(\eta)\right],
\label{varphisol}
\end{equation}
where the homogeneous solutions $\varphi_k^{1,2}$ are given by
\begin{equation}
\varphi_k^{1,2}=\frac{1}{2a} (\pi|\eta|)^{1/2}
                        H_\nu^{(1),(2)}(k\eta).
\end{equation}
Here $H_\nu^{(1)}$ and $H_\nu^{(2)}$ are Hankel functions of the
first and second kinds respectively and $\nu^2=9/4-m_{\varphi{\rm
eff}}^2/H^2$. In addition, we have from Eq.~(\ref{sigmaeq}) that
\begin{equation}
\sigma_k=\frac{1}{a} (\pi|\eta|)^{1/2} \left[c_1
H_\mu^{(1)}(k\eta)+c_2 H_\mu^{(2)}(k\eta)\right],
\end{equation}
where the constants $c_1$ and $c_2$ are subject to the
normalization condition, $|c_2|^2 -|c_1|^2=1$, and
$\mu^2=9/4-m_{\sigma{\rm eff}}^2/H^2$.

Now, the power spectrum of the perturbation $\varphi$ can be obtained by 
employing Eqs.~(\ref{noise}) and (\ref{varphisol}) as
\begin{equation}
\langle\varphi_{\vec k}(\eta)\varphi_{{\vec k}'}^*(\eta)\rangle
=\frac{2\pi^2}{k^3}\Delta^\xi_k(\eta) \delta({\vec k}-{\vec k}'),
\end{equation}
where the noise-driven power spectrum is given by
\begin{equation}
\Delta^\xi_k(\eta)
=\frac{g^4 z^2}{8\pi^4} \int_{z_i}^z dz_1
\int_{z_i}^z dz_2 {\bar\phi}(\eta_1) {\bar\phi}(\eta_2)
\frac{\sin z_{-}}{z_1 z_2 z_{-}}
\left[z_{-}^{-1}\sin\frac{2\Lambda z_{-}}{k}-1\right] F(z_1) F(z_2),
\label{pseq}
\end{equation}
where $z_{-}=z_2-z_1$, $z=k\eta$, $z_i=k\eta_i=-k/H$, $\Lambda$
is the momentum cutoff introduced in the evaluation of the ultraviolet
divergent Green's function in Eq.~(\ref{grfct}), and $F(z)$ is given by~\cite{cnoise}
\begin{equation}
F(y)=\left(1+\frac{1}{yz}\right)\sin(y-z)+ \left({1\over
y}-{1\over z}\right)\cos(y-z).
\end{equation}

\begin{figure}[th]
\centerline{\psfig{file=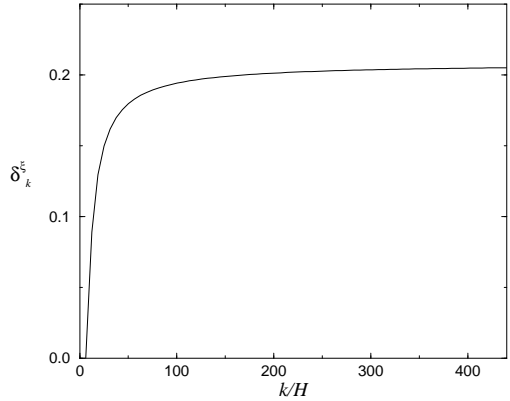,width=3.0in}}
\vspace*{8pt}
\caption{Power spectrum of the noise-driven inflaton fluctuations
$\delta^\xi_k\equiv 4\pi^2\Delta^\xi_k/g^4{\bar\phi}_0^2$.
The starting point, $k/H=2\pi$, corresponds to the $k$-mode that
leaves the horizon at the start of inflation.\protect\label{fig1}}
\end{figure}

We plot $\Delta^\xi_k(\eta)$ at the horizon-crossing time given by
$z=-2\pi$ versus $k/H$ in Fig.~\ref{fig1}. The figure shows that the
noise-driven fluctuations depend on the onset time of inflation
and approach asymptotically to a scale-invariant power spectrum
$\Delta^\xi_k\simeq 0.2g^4{\bar\phi}_0^2/(4\pi^2)$ at large $k$.
Note that if $g^4{\bar\phi}_0^2\simeq H^2$,
$\Delta^\xi_k$ will be comparable to that of the known
scale-invariant quantum fluctuation of the inflation which is
given by $\Delta^q_k=H^2/(4\pi^2)$~\cite{hawking}.

\section{Effects on the Large-Scale Structure}
\begin{figure}[th]
\centerline{\psfig{file=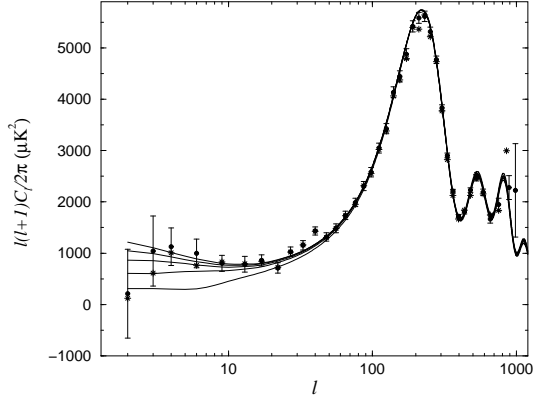,width=3.0in}}
\vspace*{8pt}
\caption{CMB anisotropy in the $\Lambda$CDM model with the
noise-driven density power spectrum $P^\xi_k$. The curves from
below represent respectively different $P^\xi_k$'s with
$k/H=200\pi$, $500\pi$, $1000\pi$, and $2000\pi$ corresponding to
the physical scale of $0.05$Mpc$^{-1}$. The top curve is the
$\Lambda$CDM model with a scale-invariant power spectrum. We
normalize all the anisotropy spectra at the first Doppler peak.
Also shown are the three-year WMAP data including error bars and 
the first year WMAP data denoted by stars.\protect\label{cl1}}
\end{figure}

In contrast to the scale-invariant density power spectrum induced by
$\Delta^q_k$, the noise-driven density power spectrum $P^\xi_k$
induced by $\Delta^\xi_k$ is blue-tilted on large scales. Assuming
a dominant $P^\xi_k$ and using the set of cosmological parameters
measured by WMAP~\cite{wmap}, we have run the CMBFAST numerical
codes~\cite{sel} to compute the CMB anisotropy power spectrum as
shown in Fig.~\ref{cl1}. We can see that the known result of the
scale-invariant power spectrum at small scales can be achieved by
properly choosing the initial time of inflation. Meanwhile, the
suppressed large-scale density perturbation can account for the
low CMB low-$l$ multipoles.

\begin{figure}[th]
\centerline{\psfig{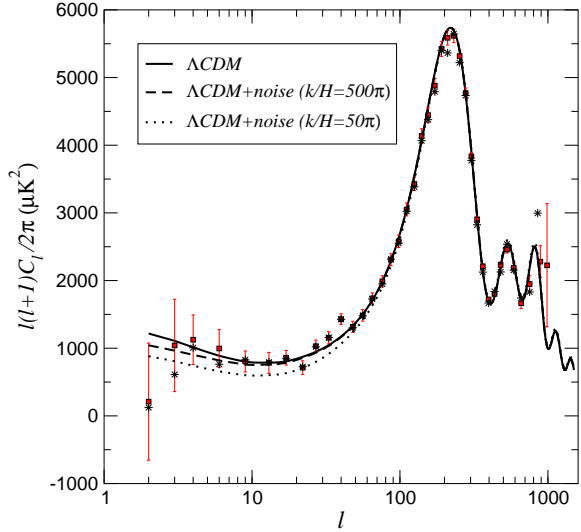}}
\vspace*{8pt}
\caption{CMB anisotropy in the $\Lambda$CDM model with the
density power spectrum $P_k=P^q_k+P^\xi_k$.
The solid curve is the $\Lambda$CDM model with a scale-invariant
$P^q_k$ induced by quantum fluctuations. The dashed and dotted curves
represent respectively noise-driven $P^\xi_k$'s with $k/H=500\pi$
and $k/H=50\pi$ corresponding to $0.05$Mpc$^{-1}$.
\protect\label{cl2}}
\end{figure}

The above results do not include the intrinsic fluctuation of inflaton. To take that into account, an additional white noise term $\xi_w$ in the free field case must be involved on the right hand side of the Langevin equation Eq.~(\ref{lange}) as
\begin{eqnarray}
\ddot{\phi}+2aH\dot{\phi}-\nabla^{2}\phi+a^{2}\left(V'(\phi)
+g^{2}\langle\sigma^{2}\rangle\phi\right)-g^{4}a^{2}{\phi}\,\times&&
\nonumber \\
\int d^{4}x' a^4(\eta') \theta(\eta-\eta')\,i\, G_{-}(x,x')
{\phi}^{2}(x')=\frac{{\phi}}{a^{2}}\xi+\xi_w. &&
\end{eqnarray}
Accordingly, the effective mass of the $\sigma$-field in Eq.~(\ref{sigmaeq}) is modified as $m_{\sigma{\rm eff}}^2=m_{\sigma}^2+g^{2}({\bar\phi}^2+ \langle\varphi_q^{2}\rangle)$, where $\langle\varphi_q^{2}\rangle$ represents the active quantum fluctuation with a scale-invariant power spectrum given by $\Delta^q_k=H^2/(4\pi^2)$~\cite{hawking}.
The total density power spectrum then is given by 
\begin{equation}
P_k=P^q_k+P^\xi_k=X({\bar\phi})\left(\Delta^q_k+\Delta^\xi_k\right),
\end{equation}
where $X({\bar\phi})$ is determined by the slow-roll kinematics. Under the circumstances, the power spectrum of the noise-driven inflaton fluctuations is not affected. However, the CMB anisotropy in the $\Lambda$CDM model are modified accordingly, as shown in Fig.~\ref{cl2} where the kinematic condition $X({\bar\phi}=\bar{\phi}_0)$ has been set to constant. 

The passive fluctuation induced by the quantum noise not only suppresses the CMB power on large scales, it also impresses on the trajectory of spectral index. If we rewrite the total density power spectrum as
\begin{equation}
P_k=P^q_k\left(1+\frac{\Delta^\xi_k}{\Delta^q_k}\right),
\end{equation}
the energy scale of inflation inferred from measurements of the density power
spectrum will be lower than the standard slow-roll prediction by a
factor $(1+\Delta^\xi_k/\Delta^q_k)^{-1/4}\simeq 0.92$. The spectal
index is then given by
\begin{equation}
n(k)-1\equiv \frac{d\ln P_k}{d\ln k}= \frac{d\ln P^q_k}{d\ln k}
+ \frac{d}{d\ln k}\ln\left(1+\frac{\Delta^\xi_k}{\Delta^q_k}\right).
\label{index}
\end{equation}
Apparently, the additional term containing
$\Delta^\xi_k$ in Eq.~(\ref{index}) offers a new dynamical source for
breaking the scale invariance. It is worthwhile to study
the overall effect of the corrections induced by the dissipation,
the static mass approximation, and the static mean field
approximation to the spectral index.
Especially, we expect a running spectral index at large $k$
due to dissipational effects on the fluctuations at late times.
Furthermore, we have made a Gaussian approximation to derive
the two-point function of the colored noise in Eq.~(\ref{noise}).
Actually, the noise-induced density fluctuations are intrinsically
non-Gaussian. So, it is interesting to go beyond
the Gaussian approximation and consider the three-point function
$\langle\xi(x_1)\xi(x_2)\xi(x_3)\rangle$,
which would directly lead to non-zero
$\langle\varphi_{\vec k_1}\varphi_{\vec k_2}\varphi_{\vec k_3}\rangle$
without invoking higher-order terms. 

\section{Conclusion}
we have proposed a new scenario in which the
inflaton is viewed as a phenomenological classical field. In this
model, the mean field drives the kinematics of inflation, while
its fluctuations are dynamically generated by the colored quantum
noise as a result of the coupling of the inflaton to a quantum
environment. This model is a better fit to the WMAP anisotropy
data than the $\Lambda$CDM model. Moreover, the observed low CMB
quadrupole may unveil the initial condition for inflation. These
results are derived by solving approximately the Langevin
equation~(\ref{lange}) of the inflaton which is coupled to a
massive quantum field. The noise term then generates a stochastic
force which drives the inflaton fluctuations. It is worth studying
the overall effect of the corrections induced by the dissipation,
the static mass approximation, and the static mean field
approximation to the spectral index of the density power spectrum.
Furthermore, the density power spectrum driven by the colored
noise may be non-Gaussian.

Although we are based on a simple inflaton-scalar model, our
results should be generic in any interacting model. We have
considered a quantum scalar field with mass of the order of the
Hubble parameter during inflation. In fact, one can consider a
scalar field with much higher mass. So it is mandatory to perform
a full analysis of the cosmological effects due to the quantum
environment in a viable interacting inflation model such as the
hybrid or $O(N)$ model.

\section*{Acknowledgments}
This paper is based on the work, astro-ph/0604292, collaborated with Chun-Hsien Wu, Kin-Wang Ng, Da-Shin Lee and Yeo-Yie Charng. The author thanks them all for many inspiring discussions during the pleasant times we worked together.

\end{document}